%% file: reduction3.tex
\documentclass[twoside]{book}
\usepackage{graphicx}           
\input ICCSsty.tex              

\newcommand{\qed}{\vrule height1.2ex width1ex depth.1ex}
\newenvironment{proof}{\par\noindent{\sc Proof}\quad}{\qed\par\smallskip}

\newtheorem{theorem}{Theorem}
\newtheorem{lemma}[theorem]{Lemma}
\newtheorem{definition}{Definition}

\newcommand\rcsinfo{ \$Source: /home/axel/paper/reduction3/RCS/reduction3.tex,v $ $ --
  \$Revision: 2.7 $
  $ -- \$Date: 2005/02/06 04:34:12 $ $ }

\shtitle{No need to blur the picture}      
\title{No need to blur the picture}
\author{Axel G. Rossberg\affil{Yokohama National University, Japan\\rossberg@ynu.ac.jp}}

\abstract{A formalism specifying efficient, ``emergent'' descriptions
  of experimental systems is developed.  It does not depend on an
  \textit{a priori} assumption of limited available data.}

\usepackage{amsmath}
\usepackage{amsfonts}
\usepackage{amssymb}
\usepackage{url}
\usepackage{psfrag}

\newcommand{\pow}{{\mathop{\mathrm{pow}}}}
\newcommand{\opt}{{\mathop{\mathrm{opt}}}}

\newcommand{\Y}{Y}

\begin{document}
\date{\tiny\rcsinfo}




\maketitle

\section{Introduction}
\sectlabel{sec:introduction}

A complex systems can become an economical problem.  Understanding its
internal machinery, describing it, and predicting its future behaviour
can be expensive.  The problem of finding simple, accurate, and
efficient descriptions is a central aspect of the work on complex
systems.  Perhaps it is \emph{the} unifying aspect of complex-systems
science.

Interestingly, this practical problem is closely related to the
philosophical problem of emergence \cite{sep-properties-emergent,%
  damper00:_editor_special_issue_emerg_proper_compl_system}.  Stated
in its weakest form, this is the question why, if the basic laws of
physics are so simple, the world around us appears to have such a rich
structure.  A partial answer that easily comes to mind is this: If we
would try to apply the basic laws every time we interpret the world
around us, it would just take too much time.  Instead we are using
other descriptions that are more efficiently.  But each applies only
to a particular part of the world, so we need many of them.  In the
language of computer science \cite{li97:_kolmog}, we are trading
computation time for description length.  Apparently, this is a good
deal.  The structure of the world as we see it is a result of solving
just the economic problem mentioned above.  We are reducing the cost
of describing the complex system ``world''.

This is only a partial answer to the problem of emergence.  Many
questions remain unanswered, such as, ``Why are there distinct parts
for which efficient descriptions exist?'' or ``Can efficient
descriptions be found systematically, and, if yes, how?''.  But it is
this partial answer that will be of interest here, for it is itself
incomplete.

Efficient, simplified descriptions are rarely perfectly precise, and
somehow a decision has to be made which information about the thing
described the description should reproduce, and which may be ignored.
The conventional strategy to proceed when arriving at this part of the
problem
(e.g. \cite{castellani02:_reduc,
  crutchfield94:_calcul_emerg,%
  rissanen89:_stoch_compl_statis_inquir}) is to presupposed that the
information regarding the thing described is incomplete anyway, and
only the available information must be reproduced.  This blurring of
the picture comes under many different names: finite samples of noisy
data, coarse graining, partitioning of the state space,
\textit{e.t.c.}.  As a result, the choice of the simplified
description becomes essentially a function of the mode of observation.
But does this correspond to the facts?  The history of science knows
many examples of simplified descriptions (and related concepts) that
have been introduced long before the things described could be
observed.  Obvious examples are descriptions in terms of
quasi-particles such as ``holes'' and ``phonons'' used in solid state
physics.  On the other hand, descriptions that are much coarser than
any reasonable limit of observation are also frequently used.  One
might just think of a description of traffic flow in terms of atomic
``cars''.

Shalizi and Moore \cite{shalizi00:_what_is_macros} suggested a
solution of this problem based on \emph{causal states}
\cite{crutchfield89:_infer_statis_compl}.  Here, a different argument
for reducing the information to be reproduced by a description is
explored.  Information regarding the thing described is dropped not
because it is unavailable, but for the sake of an efficient and simple
description.  Central to this argument is the distinction between two
kinds of descriptions: {\em models}, that produce data somehow
similar to present or future real data, and {\em
 characterizations}
that summarize some aspects of data.

Predictions about complex systems generally require both: a model that
is used for the prediction, and a characterization that specifies what
aspects of the real data the model is supposed to reproduce.
By the condition that model and characterization are \emph{both}
simple and efficient, particular choices for the information to be
retained by the descriptions are singled out.  This part of the
information is ``relevant'' for a simple reason: it can be predicted
within given cost constraints.

In the remainder of this work, it is shown that this approach can be
taken beyond hand-waving.  Formal definitions of basic notions are
introduced.  Desiderata for economic descriptions are summarized under
the notion of \emph{basic model-specifying characterizations}
(b.m.s.c.), and it is shown that nontrivial b.m.s.c.\ exist.  They are
by far not unique. The accuracy and detail of preferred descriptions
depends on the available resources, and the formalism is taking this
into account.  Results are illustrated by a minimal example.

\Fig{fig:setting}{
  \includegraphics[width=0.7\columnwidth,keepaspectratio]{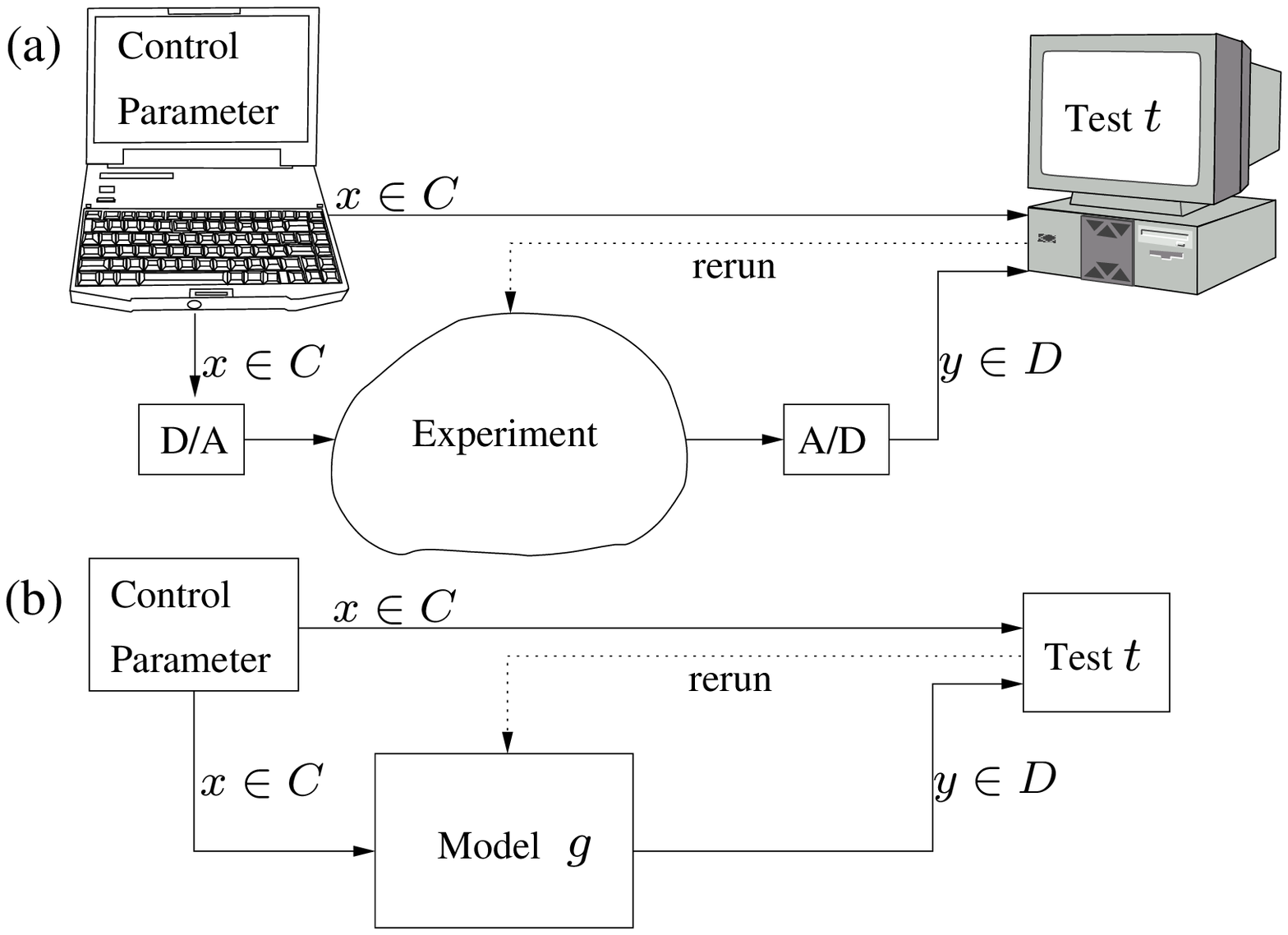}
}{(a)  Generic setup of a computer-controlled experiment.
  (b) Data flow in a test of a computational model.}

\section{The formalism}
\sectlabel{sec:formalism}

For the formal analysis, both models and characterizations are
represented by computer programs.  The complex system to be described
is represented by a computer-controlled experiment.
\fig{fig:setting} illustrates the interaction between
experimenter (the ``Control Parameter'' terminal), experiment, model,
and characterization.  A characterization of data is given by a
statement saying the data passes a certain \emph{test}; a statistical
test in general.

Throughout the theory, assume a {\df control parameter format}
$C\subset\{0,1\}^n$ and a {\df data format} $D\subset\{0,1\}^m$ to be
fixed, with $\{0,1\}^k$ denoting the set of all
binary strings of length $k$ and $n,m\in\mathbb{N}_0$.
Given a control parameter value $x \in C$ and being run, the
experiment (including the D/A and A/D conversion) produces an output
value $y\in D$.  Input and output data can be sets of numbers, images,
time-series, e.t.c..  The only major limitation is that both $C$ and
$D$ are finite sets.  The A/D conversion of the experimental output
naturally involves some loss of information.  But below it is argued
that the information passing through the A/D converter can be much
richer than the information tested for and being reproduced in the
model.  The information loss at the A/D converter is not decisive for
determining the ``emergent'' description.

In general, the complex system involved in the experiment is not
deterministic.  The experimental output $y$ is a realization of a
random variable $Y$ with values in $D$.  The experiment is assumed
reproducible in the sense that repeated runs of the experiment (with
identical $x$) yield a sequence $Y_1,Y_2,\ldots$ of statistically
independent, identically distributed (i.i.d.)\ results.

\begin{definition}
  For a given (deterministic) machine model, a {\df test} $t$ is a
  program that takes a control parameter $x\in C$ as input, runs, and
  then halts with output $0$, $1$, or $\mathbf{e}$.  When the output
  is not $\mathbf{e}$, the test can request several data samples
  before halting (``rerun'' in \fig{fig:setting}).  Then
  execution of the test is suspended until a sample $y\in D$ is
  written into a dedicated storage accessible by the test.  The number
  of samples requested can depend on the sampled $y$ but is finite for
  any sequence of successive samples.
\end{definition}
By the output $\mathbf{e}$ the tests $t$ indicates that $x$ is not
within the range of validity $C[t]:=\{x\in C|\text{output of $t$ with
  input $x$ is not $\mathbf{e}$\}}$ of the corresponding
characterization.  The outputs $1$ or $0$ indicate that the null
hypothesis (see below) is accepted or rejected by the test,
respectively.

Models are represented by \emph{generators}.
\begin{definition}
  Given a machine model, {\df generator} $g$ is a program that takes
  a control parameter $x\in C$ as input, runs, outputs data $y\in D$
  and halts.  The program has access to a source of independent,
  evenly distributed random bits in an otherwise deterministic
  machine.
\end{definition}

Now a cost functions is introduced which measures the cost involved in
running models $g$ and tests $t$, constructing and evaluating them,
and performing experiments.  We assume that this cost can be expressed
in terms of the lengths $L(t),L(g)\in\mathbb{N}_0$ of the programs $t$
and $g$, their average execution times $T(g),T(t)\in\mathbb{R}^{\ge
  0}$, and the average number $N(t)\in\mathbb{R}^{\ge 0}$ of
experimental runs required by $t$.  To be specific, define $T(\cdot)$
as the maximum of the expectation value of the runtime over all $x\in
C$ and all distributions of input data, $N(\cdot)$ analogously.  It
can be shown that $T(t)$ and $N(t)$ are always finite.  As
conventional, the number of tests or generators $q$ with $L(q)\le n$
is assumed to be finite all $n\in\mathbb{N}_0$.
\begin{definition}
  A {\df cost function} $K$ is a mapping
  $K:\mathbb{N}_0\times\mathbb{R}^{\ge 0} \to \mathbb{R}^{\ge 0}$ or
  $K:\mathbb{N}_0\times\mathbb{R}^{\ge 0}\times\mathbb{R}^{\ge
    0}\to\mathbb{R}^{\ge 0}$ that increases strictly monotonically in
  all its arguments.  The abbreviation $K(t)$ stands for
  $K[L(t),T(t),N(t)]$ if $t$ is a test and $K(g)$ stands for
  $K[L(g),T(g)]$ if $g$ is a generator.
\end{definition}
In practice, the cost of descriptions depends strongly on the
circumstances.  The theory should therefore be independent of the
particular choice of the cost function.  For this purpose, as is made
clear by Theorem~\ref{thm:minima} below, the following definition is
convenient.
\begin{definition}
  Let $p_1$ and $p_2$ be two tests or two generators.  Then the
  relations $\preceq$ ({\df always cheaper or equal}) and $\prec$
  ({\df always cheaper}) are defined by
  \begin{align}
    \eqlabel{def:cheaperorequal}
    p_1 \preceq p_2 &\stackrel{\text{def}}{\Leftrightarrow}
    \text{$L(p_1)\le L(p_2)$ and $T(p_1)\le T(p_2)$ and $N(p_1) \le N(p_2)$}
  \end{align}
  {(for generators without the last condition) and}
  \begin{align}
    \eqlabel{def:cheaper}
    p_1 \prec p_2 &\stackrel{\text{def}}{\Leftrightarrow}
    \text{$p_1\preceq p_2$ and not $p_2\preceq p_1$}.
  \end{align}
  
  A test or generator $p$ is said to be {\df $\prec$-minimal} in a
  set $P$ of tests or generators if $p\in P$ and there is no $p'\in P$
  such that $p'\prec p$.
\end{definition}
\begin{lemma}
  \label{thm:props}
  Relation $\preceq$ is transitive and reflexive, relation $\prec$ is
  transitive and antireflexive.
\end{lemma}
(Since $\preceq$ is not antisymmetric, it is not a partial order.)
The proof is standard.
\begin{lemma}
  \label{thm:order}
  For any two tests or generators $p_1$, $p_2$, and any cost function
  $K$, $p_1\prec p_2$ implies $K(p_1)<K(p_2)$.
\end{lemma}
\begin{proof}
  Assume that $p_1$ and $p_2$ are generators.  Then $L(p_1)\le L(p_2)$
  and $T(p_1)\le T(p_2)$ and either $L(p_1)<L(p_2)$ or
  $T(p_1)<T(p_2)$, since if both were equal the last part of
  condition~\eq{def:cheaper} would be violated.  Thus, using the
  strict monotony of $K$, one has either
  $K[L(p_1),T(p_1)]<K[L(p_2),T(p_1)] \le K[L(p_2),T(p_2)]$ or
  $K[L(p_1),T(p_1)]\le K[L(p_2),T(p_1)] < K[L(p_2),T(p_2)]$.  Both
  imply $K(p_1)<K(p_2)$.  For tests the proof is analogous.
\end{proof}
\begin{theorem}
  \label{thm:minima}  Let $P$ be a set of tests or generators. 
  $p\in P$ is $\prec$-minimal in $P$ if and only if there is a cost
  function $K$ that attains its minimum over $P$ at $p$.
\end{theorem}
\begin{proof}
  The ``if'' part: If some $K$ would attain its minimum over $P$ at
  $p$ but $p$ was not $\prec$-minimal, there would be a $p'\in P$ such
  that $p'\prec p$ and, by Lemma~\ref{thm:order}, $K(p')<K(p)$.  But
  this contradicts the premise.  So $p$ is $\prec$-minimal.
  
  The ``only if'' part: Assume $p$ is $\prec$-minimal in a set of
  generators $P$.  We show that there is a cost function that attains
  its minimum over $P$ at $p$ by explicit construction.
  $K(l,t):=\kappa(l,L(p))+\kappa(t,T(p))$ with $\kappa(z,z_0)=z$ for
  $z\le z_0$ and $\kappa(z,z_0)=L(p)+T(p)+z$ for $z>z_0$ does the job.
  Obviously $K$ satisfies strict monotony.  And any $p'\in P$ that
  does not have $L(p')=L(p)$ and $T(p')=T(p)$ [and hence $K(p')=K(p)$]
  must have either a larger $L$ or a larger $T$ than $p$, otherwise
  $p$ would not be $\prec$-minimal.  But then $K(p') \ge
  L(p)+T(p) = K(p)$.  So $K(p)$ is the minimum of $K$ over $P$.  For
  tests the proof is analogous.
\end{proof}
\begin{lemma}
  \label{thm:min-exists}
  Every nonempty set $P$ of tests or generators contains an element
  $p$ which is $\prec$-minimal in $P$.
\end{lemma}
\begin{proof}
  Assume that $P$ has no $\prec$-minimal element.  Then for every
  element $p\in P$ there is a $p'\in P$ such that $p'\prec p$.  Thus
  an infinite sequence of successively always-cheaper ($\prec$)
  elements of $P$ can be constructed.  Because $\prec$ is transitive
  and antireflexive, such a sequence contains each element at most
  once.  Let $q$ be the first element of such a sequence.  Since by
  definition $p\prec q$ implies $L(p)\le L(q)$, and there is only a
  finite number of programs $q$ with $L(q)\le L(p)$, the number of
  successors of $p$ cannot be infinite.  So the premise that $P$ has
  no $\prec$-minimal element is wrong for any nonempty $P$.
\end{proof}
The $\prec$-minimal element is generally not unique.  Different
$\prec$-minima minimize cost functions that give different weight to
the resources length, time, and, experimental runs used.  On the other
hand, it turns out that in practice the machine dependence of relation
$\prec$ for implementations of algorithms on different processor
models is weak.  Therefore, instead of cost functions, relation
$\prec$ is used below.

\medskip

A central element of statistical test theory
\cite{lehmann97:_testinSHORT} is the \emph{power function}.  It is
defined as the probability that the test rejects data of a given
(usually parameterized) distribution.  The goal of statistical test
theory is to find tests who's power function is below a given
significance level $\alpha$ if the null-hypothesis is satisfied, and
as large as possible otherwise.  

Denote by the {\df test function} $t_x(\{y_i\})$ the output of the
test $t$ at control parameter $x\in C[t]$ when applied to the sequence
of experimental results $\{y_i\}\in D^\infty$ (for formal simplicity,
the sequences $\{y_i\}$ are assumed infinite, even though tests use
only finite subsequences).
\begin{definition}
  For any test $t$, the {\df power} of the test function $t_x$, when applied
  to the random sequence $\{Y_i\}$ with values in $D^\infty$, is the
  probability of rejecting $\{Y_i\}$, i.e.,
  \begin{align}
    \eqlabel{def:power}
    \pow(t_x,\{Y_i\}):=\Pr\!\left[t_x(\{Y_i\})=0 \right]\quad(x\in C[t]).
  \end{align}
\end{definition}
Unlike in conventional test theory, there is no independent null
hypothesis $H_0$ here that states the distribution or the class of
distributions of $\{Y_i\}$ that is tested for.  Instead, given a test
function $t_x$, the null hypothesis, i.e., the class of distributions,
is \emph{defined} by the condition
\begin{align}
  \eqlabel{def:correct_tests}
  \pow(t_x,\{Y_i\})\le\alpha,
\end{align}
where $0<\alpha<1$ is a fixed\footnote{From $t_x$ tests for the same
  $H_0$ at other significance levels can be constructed.}
significance level.

\bigskip

Now the concepts from statistics and computer science introduced above
are combined.  Denote by $g_x$ the sequence $\{Y_i\}$ of random
outputs of generator $g$ at control parameter $x$.
\begin{definition}
  A generator $g$ is an {\df optimal generator} relative to a test
  $t$ and a power threshold $1>\gamma>\alpha$ (notation:
  $\opt_t^\gamma g$) if
  \begin{enumerate}
  \item $\pow(t_x,g_x)\le\alpha$ for all $x\in C[t]$ and
  \item for every generator $g' \prec g$ there is a $x\in C[t]$ such
    that $\pow(t_x,g_x')>\gamma$.
  \end{enumerate}
\end{definition}
This implies that $g$ is $\prec$-minimal in
$\{g'|\text{$\pow(t_x,g'_x)\le\alpha$ for all $x\in C[t]$}\}$. Hence
$g$ is, for some cost function, the minimal (-cost) model for the
property that $t$ is testing for.  Condition 2.\ can be satisfied only
for particular choices of $t$.  It requires a minimal power $\gamma$
from $t$ to distinguish the models that it characterizes from those is
does not.  Constructing tests that maximize $\gamma$ leads to results
similar to the \emph{locally most powerful tests} of statistical test
theory~\cite{lehmann97:_testinSHORT}.

For an i.i.d.\ random sequence $\{Y_i\}$ denote by $p[\{\Y_i\}]$ the
distribution function of its elements, i.e.,
$p[\{Y_i\}](y):=\Pr[Y_1=y]$ for $y\in D$.  
\begin{definition}
  Call a generator $g$ an {\df optimal implementation} with respect
  to a set $\tilde C\subset C$ if it is $\prec$-minimal in
  $\{g'|\text{$p[g'_x]\equiv p[g_x]$ for all $x\in \tilde C$}\}$ (the
  set of generators that do exactly the same).
\end{definition}
\begin{theorem}
  \label{thm:t-exists}
  For every $\tilde C\subset C$, every optimal implementation $g$ with
  respect to $\tilde C$, and every $1>\gamma>\alpha$ there is, a test
  $t$ such that $\opt_t^\gamma g$ and $C[t]=\tilde C$.
\end{theorem}
\begin{proof}
  An explicit construction of $t$ is outlined: $x\in\tilde C$ can be
  tested for by keeping a list of $\tilde C$ in $t$.  Since there is
  only a finite number of $g'\preceq g$, the test must distinguish
  $p[g_x]$ from a finite number of different distributions $p[g'_x]$
  for all $x\in\tilde C$, with power $\gamma$.  This can be
  achieved by comparing a sufficiently accurate representation of
  $p[g_x]$, stored in $t$ for all $x\in \tilde C$, with a histogram
  obtained from sufficiently many samples of $g'_x$. 
\end{proof}

\begin{definition}
  Call a pair $(t,g)$ a {\df basic model-specifying characterization}
  (b.m.s.c.)\ if $t$ is $\prec$-minimal in
  $\{t'|\text{$\opt^\gamma_{t'} g$ and $C[t]\subset
  C[t']$}\}$ for some  $1>\gamma>\alpha$.
\end{definition}
That is, for some cost function the test $t$ gives the minimal
characterization required to specify $g$ (given power threshold
$\gamma$ and range of validity $C[t]$).  Sometimes there are other
generators which are similar to $g$ but cheaper.  Then $t$ must be
very specific to characterize the particularities of $g$.  In other
cases, the output of $g$ has an essentially new, ``striking'' property
which cannot be obtained with cheaper generators.  If the property is
really ``striking'', a rather cheap and generic test $t$ is sufficient
to detect it.  \emph{Thus $t$ can ignore all other information
  contained in the output of $g$.}  Such an approximate
characterization is most likely to apply also to the data of an actual
experiment.  Then the b.m.s.c.\ $(t,g)$ provides a specific but
economic description.  After verifying the b.m.s.c.\ for some control
parameters $x\in C[t]$, approximate predictions of experimental results
for other parameters can be obtained from $g$ by the usual (though
philosophically opaque) method of induction.

A trivial b.m.s.c.\ is given by a test $t$ that always outputs $1$ and
some generator $g$ $\prec$-minimal among all generators.  But the
following makes clear that the world of b.m.s.c.\ is much richer.
\begin{theorem}
  \label{thm:bmsc-exists}
  There is, for every $\tilde C\subset C$ and every optimal
  implementation $g$ with respect to $\tilde C$, a test $t$ such that
  $(t,g)$ is a b.m.s.c.\ and $\tilde C \subset C[t]$.
\end{theorem}
\begin{proof}
  Fix some $1>\gamma>\alpha$.  By Theorem~\ref{thm:t-exists}, the set
  $S:=\{t'|\opt^\gamma_{t'} g \text{ and } \tilde C \subset C[t']\}$ is
  nonempty.  Theorem~\ref{thm:bmsc-exists} is satisfied by any $t$
  which is $\prec$-minimal in $S$.  By Lemma~\ref{thm:min-exists} such an
  element exists.
\end{proof}

\section{A simple example}
\sectlabel{sec:examples}

As a minimal, analytically traceable example, consider an experiment
without control parameters $C=\emptyset$ in which only a single bit is
measured, $D=\{0,1\}$. The probability $p$ for the cases $y=0$ to
occurs is exactly $p=0.52$ and the ``complexity'' of the systems
consists just in this nontrivial value.  With $\alpha=0.1$, the
following pair $(t,g)$ is a b.m.s.c.: A generator $g$ [with
$L(g)=52\,\mathrm{byte}$ and $T(g)=56\,\mathrm{\upsilon}$ on the {\sc mmix}
model processor \cite{knuth99:_mmixw}; the unit of time reads
``oops''] that outputs $y=0$ and $y=1$ with exactly equal probability
$p=1/2$, and a test $t$ ($L(t)=104\,\mathrm{byte}$ and
$T(t)=255\,\mathrm{\upsilon}$) that verifies if among $N=5$ samples
both $y=0$ and $y=1$ occur at least once.  This test is the cheapest
test that accepts the model $g$ ($\pow(t,\{g\})=1/16\le\alpha$) and
rejects all cheaper models, namely generators $g'$ that always output
the same value [one finds $L(g')=28\,\textrm{byte}$,
$T(g')=38\,\upsilon$, $\pow(t,\{g'\})=1>\alpha$].  But $t$ also
characterizes all experiments for which
$\pow(t,\{Y_i\})=p^N+(1-p)^N\le\alpha$, such as our case $p=0.52$,
where $\pow(t,\{Y_i\})\approx 0.064$.

There are other b.m.s.c.\ for the experiment.  For example, a
generator $g^*$ that computes a 8-bit random integer in the range
$0,...,2^8-1$, and uses it to output $y=0$ with probability
$p=133\times 2^{-8}=0.5195$ and $y=1$ otherwise
[$L(g^*)=76\,\mathrm{byte}$ and $T(g^*)=225\,\upsilon$]; and a test
$t^*$ that verifies if within 962 samples between 437 and 487 cases
$y=0$ occur [$L(t^*)=112\,\textrm{byte}$, $T(t^*)=40430\,\upsilon$].
One finds $\pow(t^*,\{g^*\})=0.099834\le\alpha=0.1$ and
$\pow(t^*,\{Y_i\})=0.099832\le\alpha$ for the experimental data.  The
next cheapest generators, which have $p=132\times 2^{-8}=0.5156$ or
$p=134\times 2^{-8}=0.5234$, and are faster because they require only
6-bit or 7-bit random numbers respectively, are rejected with a power
larger than $\gamma=0.108576>\alpha$.  A cheaper test could not
reach this $\gamma$.


One might think of $g$, $g*$, and some exact $g^{**}$ as a primitive
from of different levels of description for the same experiment.

\bibliography{/home/axel/bib/bibview}

\end{document}

%% file: ICCSsty.tex
 \makeatletter
\ifx\UNDEF\mail\def\mail{ }\else\fi
\ifx\UNDEF\prange\def\prange{0 0}\else\fi

\gdef\@empty{}
\def\Mail#1 #2 {\gdef\thecontact{#1}\gdef\theaddr{#2}}
\def\Range#1 #2 {\gdef\thefirstpage{#1}\gdef\thelastpage{#2}}
{\let\'\mail \expandafter\Mail\' }	
{\let\'\prange \expandafter\Range\' }	
 \gdef\@shtitle{\relax}
 \long\def\shtitle#1{\gdef\@shtitle{#1}}
 \long\def\author#1{\gdef\@author{#1}}
 \def\affil#1{\par\noindent{\rm#1\par}}
 \gdef\@abstract{}
 \long\def\abstract#1{\gdef\@abstract{#1}}

 \renewcommand{\@evenhead}{\thepage\qquad\qquad\@shtitle\hfil}
 \renewcommand{\@oddhead}{\hfil\@shtitle\qquad\qquad\thepage}

 \def\maketitle{\thispagestyle{empty}\chapter{\@title}}
 \renewcommand\chapter{\if@openright\cleardoublepage\else\clearpage\fi
                    \thispagestyle{empty}%
                    \global\@topnum\z@
                    \@afterindentfalse
                    \secdef\@chapter\@schapter}
 \def\@makechapterhead#1{%
  \vspace*{50\p@}%
  {\parindent \z@ \raggedleft \normalfont
    \ifnum \c@secnumdepth >\m@ne
      \if@mainmatter
        \huge \@chapapp{} \thechapter
        \par\nobreak
        \vskip 20\p@
      \fi
    \fi
    \interlinepenalty\@M
    \Huge \bfseries #1\par\nobreak
    \vskip.25in
    \large\bfseries\@author\par\nobreak
    \vskip 40\p@}
    \ifx\@abstract\@empty\else{\small\@abstract\par\vskip20\p@}\fi
  }

 \let\df\bf

\DeclareRobustCommand\em
        {\@nomath\em \ifdim \fontdimen\@ne\font >\z@
                       \upshape \else \slshape \fi}

\def\@begintheorem#1#2{\sl \trivlist \item[\hskip \labelsep{\bf #1\ #2}]}
\def\@opargbegintheorem#1#2#3{\sl \trivlist
     \item[\hskip \labelsep{\bf #1\ #2\ (#3)}]}

 \newcommand{\eq}[1]{(\ref{eq:#1})}	
 \newcommand{\fig}[1]{Fig.~\ref{fig:#1}}

 \newcommand{\sectlabel}[1]{\label{sect:#1}}
 \newcommand{\eqlabel}[1]{\label{eq:#1}}
 \newcommand{\figlabel}[1]{\label{fig:#1}}

  \def\@arabic#1{\number #1} 

\long\def\@makecaption#1#2{
	\vskip\abovecaptionskip
	\sbox\@tempboxa{{\small {\bf #1}: #2}}%
	\ifdim\wd\@tempboxa>\hsize
	    {\small {\bf #1}: #2\par}
	\else
	   \global\@minipagefalse
	   \hbox to\hsize{\hfil\box\@tempboxa\hfil}
	\fi
	\vskip \belowcaptionskip}

\def\figstrut#1{\hbox to\linewidth{\vrule height#1\hfill}}

\newcommand{\Fig}[4][!htb]{
\begin{figure}[#1]
 \centering\leavevmode#3%
 \caption{#4}
 \figlabel{#2}
\end{figure}                 }

\makeatother